\documentclass[
	aps
	,a4paper
	,notitlepage
	,pra
	,eqsecnum
	,showpacs
	,floatfix
	,twocolumn
	,merge
	,elide
			]{revtex4-1}
\usepackage{amsmath}
\usepackage{amsfonts}		
\usepackage{graphicx}		
\usepackage{tikz}			
\usepackage{fourier}		
\usepackage{amssymb}
\usepackage[%
	unicode=true
	,colorlinks=true
	,linkcolor=blue
	,anchorcolor=blue
	,breaklinks=true
	,citecolor=blue
	,filecolor=cyan
	,menucolor=blue
	,breaklinks=true
	,urlcolor=blue
			]{hyperref}

\begin{document}
	\title{On the position representation of mechanical power, force and torque operators}
	\author{M.~Georgiev}
	\email{mgeorgiev@issp.bas.bg}
	\affiliation{Institute of Solid State Physics, Bulgarian Academy of Sciences,
		Tsarigradsko Chauss\'ee 72, 1784 Sofia, Bulgaria}
	
	\date{\today}
	\begin{abstract}
		Quantizing the transfer of energy and momentum between interacting 
		particles, we obtain a quantum impulse equation and relations that the 
		corresponding mechanical power, force and torque satisfy. 
		In addition to the energy-frequency and momentum-wavelength 
		relations, we introduce the power-frequency and force-wavelength 
		analogs, respectively.
		Further, we obtain an operator representation for the mechanical power 
		and impact force in the position space and discuss their 
		correspondence with the relevant momentum operator.
		The position representation of the torque operator and its relation to 
		the orbital angular momentum operator is also considered.
		The results are grounded by the presence of a constant that appears as 
		fundamental as the Planck's constant to all obtained relations.
	\end{abstract}
	\maketitle
	

\section{Introduction}
	
	Quantum mechanics is considered as one of the greatest achievements in 
	theoretical physics \cite{mehra_historical_2001,band_quantum_2013,dirac_2009}.  
	Describing phenomena governed by principles beyond those defining the 
	macroscopic world, it would always raise discussions on a possible 
	interdependence with classical mechanics
	\cite{gruber_transition_1976,kun_quantum-classical_2000,ghose_continuous_2002,rabinowitz_is_2008,huang_correspondence_2008,page_quantum_2009,strange_semiclassical_2014,radonjic_ehrenfest_2014,blaszak_quantum_2019,hawton_maxwell_2019}. 
	The foundations of quantum mechanics are well established and verified in time 
	\cite{scully_1989,wolf_testing_1990,lamoreaux_review_1992,cinelli_2005,shadbolt_testing_2014,restuccia_photon_2019}. 
	In that respect, the quantum theory appears as an invaluable 
	tool for interpreting experimental findings from the nano-scale world and thus being 
	a ground for vast number of researches accounting for different atomic and subatomic 
	processes. Some prominent examples in the filed of condensed matter physics are the 
	photon-electron interactions  
	\cite{breidbach_2005,timmers_ultrafast_2012,ossiander_attosecond_2017},
	the electrons' interactions that underpin the magnetic properties of matter 
	\cite{defotis_1990,regnault_2002,han_2004,ummethum_2012,panja_2017,hanninen_2018}
	and different transport phenomena 
	\cite{soodchomshom_2013,he_2014,dapor_transport_2017,gehring_single-molecule_2019,paulsen_organic_2019,zhang_insights_2019}.
	Furthermore, determining the time response of the electrons \cite{hassan_optical_2016} 
	and the time that it takes for quantum jumps to occur \cite{minev_catch_2019}
	are really fascinating and promising achievements of quantum physics nowadays.
	From the perspective of present technological progress, the possibility of 
	observing and controlling quantum effects such as tunneling 
	\cite{khairutdinov_electron_1989,gunther_quantum_1995,gatteschi_quantum_2003}
	and entanglement 
	\cite{ye_quantum_2008,horodecki_quantum_2009,togan_quantum_2010,friis_entanglement_2019,shin_bell_2019}  
	promises a number of future applications in the field of logical devices 
	\cite{lee_mtj_2007,dsouza_experimental_2016,xu_handbook_2016,manipatruni_scalable_2019,wang_integrated_2019,coronado_molecular_2019}. 	
	
	Whether one studies problems in the field of fundamental or applied quantum 
	mechanics the definition of the Hamiltonian and momentum of all constituent 
	units of a quantum system is a key point in understanding the 
	corresponding dynamics. 
	On the other hand, the knowledge about all processes of energy and 
	momentum transfer and the rate at which these processes take place over 
	time may strongly enrich the study. Understanding the dynamics of energy and 
	momentum transfer requires additional investigation about the observables 
	such as the mechanical power that a certain interaction generate and the 
	applied forces. 
	In that respect, it is worth mention that the existence of Hermitian 
	linear operators representing these dynamical observables in the 
	coordinate space is unexplored question. Respectively, the impact of such 
	representations in all fields of quantum physics is still obscure.

	The present article discuss the quantization of energy and momentum transfer 
	processes between interacting particles within the standard quantum mechanic.
	Accordingly, we report a possible lower limit for the rate at which the these 
	processes take place over time and thus fundamental relations that the 
	relevant dynamical quantities such as the mechanical power, force and 
	torque satisfy. 
	In particular, we represent the Planck--Einstein and de Broglie relations 
	in terms of the mechanical power and force, respectively.
	We demonstrate that the impulse-momentum theorem for quantum-scale objects 
	is grounded by an impulse equation that has no classical analogue.
	Further, we introduce the position representation of the 
	operators associated to the respective power, force and torque and 
	establish the commutation relations these observables obey. The 
	uncertainty in mechanical power and force is also studied.
	A key role within all results plays a constant that has a unit measure of 
	energy. It emerges naturally in the theory and has as fundamental part in 
	all equations as the Planck's constant. The Planck's constant, however, 
	remains unique in terms of the study discussed in Refs. 
	\cite{fischbach_new_1991,lamoreaux_review_1992,battaglia_uniqueness_1993}. 
	
	The rest of this paper is organized as follows: 
	In Section \ref{sec:general} we briefly review the fundamental approaches and 
	related equations used as a ground in the present research;
	In Section \ref{sec:pft} we introduce the main equation and all 
	originating relations;  
	The results are further discussed in Section \ref{sec:exanple} within an 
	example of a system in which there is an absence of non-contact forces;  
	Section \ref{sec:conclusion} includes summary of the obtained results.


\section{Generalities}\label{sec:general}

\subsection{Classical observables}
	
	The Lagrangian and Hamiltonian mechanics are among the most successfully applied
	approaches when studying the dynamics of an arbitrary mechanical system. 
	Although their consideration usually includes generalized variables, henceforth,
	with respect to the objective to be followed, all functions and relations are 
	given in terms of Cartesian coordinates.
	
	For an isolated system consisting of a collection of point particles, the 
	Lagrangian $L=L(\boldsymbol{r}_i,\dot{\boldsymbol{r}}_i)$, 
	where $\boldsymbol{r}_i=(\boldsymbol{r}_1,\boldsymbol{r}_2,\ldots)$ are the 
	position vectors of the particles and 
	$\dot{\boldsymbol{r}}_i=(\dot{\boldsymbol{r}}_1,\dot{\boldsymbol{r}}_2,\ldots)$ 
	are the respective velocities, collects all the knowledge needed to determine 
	how the considered system evolves over time. 
	Respectively, predicting the particles configuration 
	at any given moment of time, one has to solve the Euler--Lagrange equations.
	On the other hand, knowing the relevant momenta 
	$\boldsymbol{p}_i=(\boldsymbol{p}_1,\boldsymbol{p}_2,\ldots)$, 
	the evolution in time can be studied with the aid of 
	the Hamiltonian mechanics by using the Hamiltonian
	$H=H(\boldsymbol{r}_i,\boldsymbol{p}_i)$ 
	and solving the Hamiltonian equations. Both, the Lagrangian and Hamiltonian 
	approaches are related such that one can 
	determine the total energy of the system by accounting for the relation
	\begin{equation}\label{eq:H}
		H=\sum_{i}\dot{\boldsymbol{r}}_i\cdot\boldsymbol{p}_i-L.
	\end{equation}
	
	Although knowing the total energy is usually sufficient in determining 
	the dynamical state of a mechanical system,
	studying the rate at which the transfer of energy and momentum between the 
	interacting particles takes place with time may strongly enrich the study. 
	To this end, in addition to \eqref{eq:H}, one has to account for the mechanical 
	power $P$ and the respective net forces 
	$\boldsymbol{F}_i=\mathrm{d}_t\boldsymbol{\partial}_{\dot{\boldsymbol{r}}_i}L$,
	 satisfying the relation
	\begin{equation}\label{eq:P}
		P=\sum_{i}\dot{\boldsymbol{r}}_i\cdot\boldsymbol{F}_i-W,
	\end{equation}
	where $W\equiv\dot{H}$ gives the rate at which the system's energy 
	changes with time. For closed systems $W$ is a zero function and it is
	included in equation \eqref{eq:P} for the sake of clarity. 
	Further, the impulse associated to the $i$-th particle reads
	\begin{equation}\label{eq:ImpulseClassic}
		\boldsymbol{J}_i=\int_{\mathbb{T}_i}\boldsymbol{F}_i\mathrm{d}t,
	\end{equation}
	where the domain $\mathbb{T}_i$ defines the time interval whithin 
	the $i$-th force acts. In the processes of elastic collision, for all $i$, 
	the impulse is distributed over 
	an infinitesimal short time-interval $\lambda_i\in\mathbb{T}_i$. Accordingly, 
	$\boldsymbol{F}_i$ represents an impact force given by
	$\boldsymbol{F}_i=\boldsymbol{J}_i\delta(t-\lambda_i)$.

\subsection{Quantum observables}
	
	In quantum mechanics,
	the information about the dynamics of a multi particle system, with position 
	vectors $\boldsymbol{r}_i=(\boldsymbol{r}_1,\boldsymbol{r}_2,\ldots)$,
	is embodied in a state function  
	$\Psi=\Psi(\boldsymbol{r}_i,t)$, where $||\Psi||^2=1$. 
	Each state function is related to the probability of observing the system
	in a particular dynamical state characterized by a number of physical quantities 
	with a given uncertainty in their values.
	Satisfying the Schrodinger equation 
	\begin{equation}\label{eq:Schrodinger}
		\mathrm{i}\hbar\partial_t\Psi(\boldsymbol{r}_i,t)=
		\hat{H}\Psi(\boldsymbol{r}_i,t),
	\end{equation}
	the explicit representation of $\Psi(\boldsymbol{r}_i,t)$ is related to 
	the Hamiltonian's representation, which in turn depends on the introduced 
	interactions and hence the way the system is being manipulated, i.e. observed. 
	Respectively, depending on how one interacts with the system's constituents, 
	the latter may demonstrate either 
	their wave-like or point-like character. 
	
	The wave-like character of the collective point-like particles' dynamics 
	described by \eqref{eq:H} and \eqref{eq:P}
	become apparent on the very frontier between the classical and quantum 
	theoretical approaches of representing observables. 
	One prominent example that represents 
	the wave-particle duality concept is a system composed of free 
	particles. In this case, the representation of the Hamiltonian in 
	\eqref{eq:Schrodinger} maps its classical counterpart in \eqref{eq:H} 
	accordingly. Thus, we have
	\begin{equation}\label{eq:DualHp}
		\hat{H}\Psi_{p}(\boldsymbol{r}_i,t)=
		\sum_{i}\dot{\boldsymbol{r}}_i\cdot
		\hat{\boldsymbol{p}}_i\Psi_{p}(\boldsymbol{r}_i,t)
		-\hat{L}\Psi_{p}(\boldsymbol{r}_i,t),
	\end{equation}
	where in the absence of spin degrees of freedom, the state function 
	$\Psi_{p}(\boldsymbol{r}_i,t)$ is represented as a product of plane waves 
	each with phase given by
	\begin{equation}\label{eq:PhaseEp}
		\phi_i=
		\hbar^{-1}\big(\boldsymbol{p}_i\cdot\boldsymbol{r}_i-E_it\big),
	\end{equation}
	$E_i$--denotes the $i$-th particle's kinetic energy. 
	Respectively, we have
	\begin{equation}\label{eq:PartialHp}
		\begin{array}{ll}
		\displaystyle
			\hat{H}\Psi_{p}(\boldsymbol{r}_i,t)=
			\sum_{i}E_i\Psi_{p}(\boldsymbol{r}_i,t),
		&
			\hat{\boldsymbol{p}}_i\Psi_{p}(\boldsymbol{r}_i,t)=
			\boldsymbol{p}_i\Psi_{p}(\boldsymbol{r}_i,t),
		\\ [0.3cm]
			\hat{L}\Psi_{p}(\boldsymbol{r}_i,t)=
			L\Psi_{p}(\boldsymbol{r}_i,t),
		\end{array}
	\end{equation}
	where Lagrangian $L$ and the associated operator are a functions only of a 
	kinetic terms and take into account both relativistic and non-relativistic 
	cases. As an example, for a system composed 
	of massless particles one has $\hat{L}\Psi_{p}(\boldsymbol{r}_i,t)=0$ 
	and $|\dot{\boldsymbol{r}}_i|=c$ for all $i$, 
	where $c$ is the light speed in vacuum.
	
	The phase associated to the $i$-th particle is further given by
	\begin{equation}\label{eq:Phase}
		\phi_i=
		\boldsymbol{k}_i\cdot\boldsymbol{r}_i-\omega_i t,
	\end{equation}
	where $\omega_i$ and $\boldsymbol{k}_i$ are the corresponding angular 
	frequency and wave vector, respectively.
	Both \eqref{eq:PhaseEp} and \eqref{eq:Phase} satisfy equation 
	\eqref{eq:DualHp} simultaneously and stand as a ground for the 
	Planck--Einstein and de Broglie relations given by
	\begin{equation}\label{eq:PlanckRelation}
		E_i=\hbar\omega_i,
	\qquad
		\boldsymbol{p}_i=
		\hbar\boldsymbol{k}_i,
	\end{equation} 
	respectively, that marked the origin of quantum theory.
	It is essential to emphasize that since $\Psi_{p}(\boldsymbol{r}_i,t)$ 
	represents a plane wave, the transformation of equation \eqref{eq:DualHp} 
	into \eqref{eq:H} holds under no semi-classical approximation.


\section{Mechanical power, force and torque as a quantum observables}\label{sec:pft}

	In addition to the 
	representation of energy, momentum and orbital angular momentum in 
	quantum mechanics, equation \eqref{eq:P} can be used as a ground for deriving 
	the relevant representations of mechanical power, force and torque.
	Since the power-force relation in \eqref{eq:P} doesn't depend on the 
	explicit representation of the Lagrangian in \eqref{eq:H}, its 
	quantum mechanical counterpart will be adequate to any type of interactions. 
	Therefore, for an arbitrary Hamiltonian representation in 
	\eqref{eq:Schrodinger}, the quantum mechanical analogue of \eqref{eq:P}
	reads
	\begin{equation}\label{eq:DualPF}
		\hat{P}\Psi_{F}(\boldsymbol{r}_i,t)=
		\sum_{i}\dot{\boldsymbol{r}}_i\cdot
		\hat{\boldsymbol{F}}_i\Psi_{F}(\boldsymbol{r}_i,t)
		-\hat{W}\Psi_{F}(\boldsymbol{r}_i,t),
	\end{equation}
	where the state function satisfies the following equations
	\begin{equation}\label{eq:PartialPF}
		\begin{array}{ll}
		\displaystyle
			\hat{P}\Psi_{F}(\boldsymbol{r}_i,t)=
			\sum_{i}P_i\Psi_{F}(\boldsymbol{r}_i,t),
		&
			\hat{\boldsymbol{F}}_i\Psi_{F}(\boldsymbol{r}_i,t)=
			\boldsymbol{F}_i\Psi_{F}(\boldsymbol{r}_i,t),
		\\ [0.3cm]
			\hat{W}\Psi_{F}(\boldsymbol{r}_i,t)=
			W\Psi_{F}(\boldsymbol{r}_i,t).
		\end{array}
	\end{equation}	
	Hence, $\Psi_{F}(\boldsymbol{r}_i,t)$ is given as a direct product of
	plane waves, such that the $i$-th one is characterized by the phase
	\begin{equation}\label{eq:PhaseFP}
		\varphi_i=\varepsilon^{-1}
		\big(\boldsymbol{F}_i\cdot\boldsymbol{r}_i-P_it\big),
	\end{equation}
	where $\varepsilon$ is a constant that has the unit measure 
	of energy. From \eqref{eq:DualPF} follows that, for any system, 
	in addition to the
	eigenstates in \eqref{eq:Schrodinger},
	there exists a state function $\Psi_{F}(\boldsymbol{r}_i,t)$ 
	associating to each individual process of energy and momentum transfer
	a quanta described by a plane wave. Accordingly, the energy and momentum 
	of the $i$-th quanta should be equal to the change in kinetic energy 
	$\delta E_i$ and momentum $\delta\boldsymbol{p}_i$ of the $i$-th 
	particle within the time-interval that the applied forces act.
	Therefore, for all $i$, the corresponding quanta has energy
	$\delta E_i\equiv\mathcal{E}_i=\hbar\varpi_i$ and momentum
	$\delta\boldsymbol{p}_i\equiv\boldsymbol{\rho}_i=\hbar\boldsymbol{\kappa}_i$, 
	where $\varpi_i=\delta\omega_i$ and 
	$\boldsymbol{\kappa}_i=\delta\boldsymbol{k}_i$ represent the 
	change in associated to the $i$-th particle angular 
	frequency and wave vector, respectively.
	Therefore, the phase in \eqref{eq:PhaseFP} is further given by
	\begin{equation}\label{eq:PhaseFP_2}
		\varphi_i=\hbar^{-1}
		\big(\boldsymbol{\rho}_i\cdot\boldsymbol{r}_i-\mathcal{E}_it\big)
	\end{equation}
	and hence,
	\begin{equation}\label{eq:PhaseFP_3}
		\varphi_i=
		\big(\boldsymbol{\kappa}_i\cdot\boldsymbol{r}_i-\varpi_it\big).
	\end{equation}
	As the total energy of the system is conserved the change in total momentum
	is zero 
	$\sum_{i}\boldsymbol{\rho}_i=0$ and the sum $\sum_{i}\mathcal{E}_i$ 
	equals the change in total kinetic energy.

\subsection{Power-frequency and force-momentum relations}\label{sec:wk}
	
	Representing the de Broglie hypothesis,
	equation \eqref{eq:DualPF} leads to an additional Planck--Einstein relation. 
	Taking into account \eqref{eq:PhaseFP} and \eqref{eq:PhaseFP_3}, for all $i$,
	we obtain two fundamental relations representing the mechanical 
	power $P_i$ and force $\boldsymbol{F}_i$ 
	as a function of the frequency $\varpi_i$ and the vector
	$\boldsymbol{\kappa}_i$, respectively. 
	Thus, we have
	\begin{equation}\label{eq:PErelation}
		P_i=\varepsilon\varpi_i, 
	\qquad 
		\boldsymbol{F}_i=\varepsilon\boldsymbol{\kappa}_i.
	\end{equation}
	The last relations show that the greater the frequency $\varpi_i$ associated to 
	the $i$-th quanta the greater the mechanical power generated in the process of 
	energy transfer. 
	Accordingly, the greater the change in momentum 
	$\delta\boldsymbol{p}_i=\hbar\boldsymbol{\kappa}_i$ the greater the force 
	acting on the $i$-th particle. 
	
	Equations \eqref{eq:PErelation} lead to alternative with respect to the 
	classical dynamics force-momentum relation and formulation of the mechanical 
	power. 
	Taking into account relations 
	\eqref{eq:PhaseFP_2} and \eqref{eq:PErelation}, we get
	\begin{equation}\label{eq:PF}
		P_i=\frac{\varepsilon}{\hbar} \delta E_i,
	\qquad
		\boldsymbol{F}_i=\frac{\varepsilon}{\hbar}\delta\boldsymbol{p}_i,
	\end{equation}
	where the constant $\hbar/\varepsilon$ has a unit measure of time.
	In contrast to the classical dynamics and the Ehrenfest's theorem, 
	relations \eqref{eq:PF} give a fundamental limit for the rate at 
	which the processes of energy and momentum transfer may happened 
	under the discussed conditions.
	In other words, on a quantum level the time-interval for   
	transfer energy and momentum when particles interact, appears 
	as a fundamental constant equal to $\hbar/\varepsilon$. 
	
	We would like to point out that, in the case the $i$-th particle's 
	momentum changes $N\gg N_{A}$ times,
	where $N_A$ is the Avogadro constant, the force-momentum relation can be
	rewritten according to the substitutions 
	$\boldsymbol{\rho}_i\mathrm{d} N\to\mathrm{d}\boldsymbol{p}_i$ and 
	$(\hbar/\varepsilon)\mathrm{d} N \to\mathrm{d}t$.

\subsection{Impulse-momentum theorem and torque}

	The power and force given in \eqref{eq:PF} depend 
	only on the change in particle's energy and momentum, respectively.
	As a result, for all $i$, the impulse equation associated to the 
	relation on the right hand side in \eqref{eq:PF} reads
	\begin{equation}\label{eq:ImpulseQuantum}
		\boldsymbol{J}_i=
		\frac{\hbar}{\varepsilon}\boldsymbol{F}_i.
	\end{equation}	
	The last equation represents the Impulse-momentum theorem within the 
	concept of wave-particle duality. 
	In contrast to \eqref{eq:ImpulseClassic}, the impulse in
	\eqref{eq:ImpulseQuantum} is defined for a discrete
	time-interval $\hbar/\varepsilon$.
	
	In addition to relations \eqref{eq:PF} we can further obtain an expression 
	for the torque applied to the $i$-th particle. 
	Denoting its orbital angular momentum by $\boldsymbol{l}_i$,
	and the applied torque by $\boldsymbol{\tau}_i$, from the right hand 
	side of equation \eqref{eq:PF} we have
	\begin{equation}\label{eq:Torque}
		\boldsymbol{\tau}_i=
		\frac{\varepsilon}{\hbar}\delta\boldsymbol{l}_i.
	\end{equation}
	As the transfer of a momentum between particles is restricted in time, 
	the time that it takes for the relevant orbital angular momentum to change
	is also a constant equals $\hbar/\varepsilon$. 
	We would like to point out that the total orbital angular momentum is conserved
	and hence with respect to \eqref{eq:PhaseFP_2}
	the orbital angular momentum of the associated quanta reads 
	$\mathbf{l}_i\equiv\delta\boldsymbol{l}_i$.


\section{Impact force: Zero initial or final momentum}\label{sec:exanple}	

	In reference to the system described by \eqref{eq:DualHp},
	for all $i$, the observables $\boldsymbol{F}_i$ in \eqref{eq:PErelation}
	represent impact forces occurring due to incident collisions
	between the constituent particles.  
	Therefore, interacting once the $i$-th particle's initial energy and momentum change. 
	In that respect,
	suggesting a zero initial or final momentum, depending on whether the 
	particle starts form rest or transfer all of its momentum, respectively,
	equations \eqref{eq:PErelation} reduce to
	\begin{equation}\label{eq:PErelation_2}
		P_i=\varepsilon\omega_i, 
	\qquad 
		\boldsymbol{F}_i=\varepsilon\boldsymbol{k}_i.
	\end{equation}
	Moreover, equations \eqref{eq:PF} and \eqref{eq:Torque} read
	\begin{equation}\label{eq:PF_2}
		P_i=\frac{\varepsilon}{\hbar} E_i,
	\qquad
		\boldsymbol{F}_i=\frac{\varepsilon}{\hbar}\boldsymbol{p}_i,
	\qquad
		\boldsymbol{\tau}_i=\frac{\varepsilon}{\hbar}\boldsymbol{l}_i.
	\end{equation}	
	One then may ask is there any relation that for all $i$, $P_i$ satisfies 
	in the case of zero momentum.
	Answering that question, we 
	rewrite the representations \eqref{eq:PhaseEp}, \eqref{eq:Phase} and \eqref{eq:PhaseFP} 
	in terms of the four-momentum, four-wavevector and four-force, respectively.
	Accordingly, the force-momentum relation in \eqref{eq:PF_2} can be rewritten in 
	Minkowski space, where the contravariant four-momentum is given by 
	$p^\mu_i=(E_i/c,\boldsymbol{p}_i)$. The corresponding relativistic relation then 
	reads
	\begin{equation}\label{eq:FourForce}
		\mathcal{F}^\mu_i=
		\frac{\varepsilon}{\hbar}p^\mu_i,
	\end{equation}
	where $\mathcal{F}^\mu_i=(P_i/c,\boldsymbol{F}_i)$ is the respective contravariant 
	four-force. With the aid of equation \eqref{eq:FourForce} we obtain expression for 
	the mechanical power that the $i$-th particle with rest mass $m_i$ may apply interacting 
	with other constituents form the considered system,
	\begin{equation}\label{eq:RelatP}
		P_i=\sqrt{F^2_ic^2+\frac{\varepsilon^2}{\hbar^2}m^2_ic^4},
	\end{equation}
	where $F_i=|\boldsymbol{F}_i|$.
	Hence, form \eqref{eq:RelatP} follows that for the $i$-th particle being at rest, 
	$P_i$ will depends only on the particle's rest mass $m_i$.
	For example, when that condition is satisfied, we have
	\begin{equation*}
		P_i=\frac{\varepsilon c^2}{\hbar}m_i.
	\end{equation*}
	The last relation gives the exact amount of mechanical power that will be generated 
	if in the process of interaction the $i$-th particle loses all of its rest mass within
	the time-interval $\hbar/\varepsilon$.
	The existence of such fundamental limit is due to the presence of the constant
	$\varepsilon$ according to which relations \eqref{eq:PF} 
	and hence \eqref{eq:PF_2} are defined.

\subsection{Time evolution and position representation}\label{sec:xyz}
	
	The energy and momentum transfer in a physical system is the main indication 
	for its evolution. 
	Since the power is an observable that accounts for such a 
	processes in addition to the Hamiltonian of a quantum mechanical system, 
	one can rely on the respective power operator given on the left hand side in
	equation \eqref{eq:PartialPF}.
	Therefore, in the framework of the considered case described by equations 
	\eqref{eq:PF_2}, in complement to the Schrodinger equation 
	\eqref{eq:Schrodinger}, we have
	\begin{equation}\label{eq:PartialP}
		\mathrm{i}\varepsilon\partial_t\Psi(\boldsymbol{r}_i,t)=
		\hat{P}\Psi(\boldsymbol{r}_i,t).
	\end{equation}
	The operator $\hat{P}$ commute with the respective Hamiltonian $\hat{H}$ such that
	for $\hat{H}\Psi(\boldsymbol{r}_i,t)=\sum_{i}E_i\Psi(\boldsymbol{r}_i,t)$ 
	and all $i$, one obtains the relation on the left hand side in \eqref{eq:PF_2}.
	Nevertheless, $P$ has to be treated in accordance to the problem one is solving.
	For example, let the $i$-th particle is at rest and represents an independent 
	quantum harmonic oscillator with frequency $\omega_i$. Further, consider a 
	hypothetical collision in which the $i$-th particle transfers all of its energy. 
	Then, the maximum mechanical power that can be generated within the 
	considered interaction is given by
	\begin{equation}
		P_{n_i}=\varepsilon\omega_i\left(n_i+\frac12\right),
	\qquad
		n_i\in\mathbb{N}_0.
	\end{equation}
	
	On the other hand, in the case of \eqref{eq:PF_2}, the position 
	representation of $\hat{P}_i$ reads
	\begin{equation*}
		\hat{P}_i\equiv-\frac{\hbar\varepsilon}{2m_i}\Delta_i.
	\end{equation*}
	
	For $\mathbb{K}=\{x,y,z\}$, the net force given in \eqref{eq:PF_2} is 
	associated to the three component operator 
	$\hat{\boldsymbol{F}_i}=\big(\hat{F}^{\nu}_i\big)_{\nu\in\mathbb{K}}$ that 
	has the following position representation
	\begin{equation}\label{eq:PartialF}
		\hat{\boldsymbol{F}}_i\equiv-\mathrm{i}\varepsilon\nabla_i.
	\end{equation}
	The components of the operator in \eqref{eq:PartialF} obey the commutation 
	relation
	\begin{equation*}
		\left[\hat{\beta}^{}_i,\hat{F}^{\nu}_i\right]=
		\mathrm{i}\varepsilon \delta_{\beta\nu},
	\qquad
		\beta,\nu\in\mathbb{K}
	\end{equation*}
	and commute with each one component of the relevant momentum operator.
	We would like to point out that in the considered case the kinetic energy 
	of the system is conserved. Thus, one can includes only potential energy 
	operators that commute with each particle's momentum operator.
	In that respect, it is worth giving the time derivative of the force operator,
	\begin{equation}\label{eq:FpTime}
		\mathrm{d}_t\hat{\boldsymbol{F}}_i=\frac{\varepsilon}{\hbar}
		\mathrm{d}_t\hat{\boldsymbol{p}}_i,
	\end{equation}
	where taking into account the Ehrenfest's theorem we get
	\begin{equation*}
		\mathrm{d}_t\langle\hat{F}^{\nu}_i\rangle=
		-\frac{\mathrm{i}\varepsilon}{\hbar^2}
		\left\langle\left[\hat{p}^{\nu}_i,\hat{H}\right]\right\rangle
		+ \left\langle\partial_t\hat{F}^{\nu}_i\right\rangle,
	\qquad
		{\nu}\in\mathbb{K}
	\end{equation*}
	and
	\begin{equation*}
		\mathrm{d}_t\langle\hat{p}^{\nu}_i\rangle=
		-\frac{\mathrm{i}}{\varepsilon}
		\left\langle\left[\hat{F}^{\nu}_i,\hat{H}\right]\right\rangle
		+\left\langle\partial_t\hat{p}^{\nu}_i\right\rangle.
	\end{equation*}
	Hence, $\mathrm{d}_t\langle\hat{\boldsymbol{F}_i}\rangle=0$, since the 
	system exhibits only a random elastic collisions of non-potential interactions.
	
	The observable $\boldsymbol{\tau}_i$ gives the acquired 
	orbital angular momentum of the $i$-th particle due to the applied impact 
	force $\boldsymbol{F}_i$.
	It is represented by a three component operator 
	$\hat{\boldsymbol{\tau}}_i=\big(\hat{\tau}^{\nu}_i\big)_{{\nu}\in\mathbb{K}}$
	and similar to the orbital and spin angular momenta, its components satisfy 
	the algebra 
	\begin{equation}\label{eq:algebra}
		\big[\hat{\tau}^{\gamma}_i,\hat{\tau}^{\beta}_i\big]= 
		\mathrm{i}\varepsilon \, \epsilon_{{\gamma}{\beta}{\nu}} \hat{\tau}^{\nu}_i,
	\qquad
		{\gamma},{\beta},{\nu}\in\mathbb{K}.		
	\end{equation}
	In coordinate space the torque has the following representation
	\begin{equation*}
		\hat{\boldsymbol{\tau}}_i\equiv
		-\mathrm{i}\varepsilon\big(\hat{\boldsymbol{r}}_i\times\nabla_i\big),
	\end{equation*} 
	where $\hat{\boldsymbol{r}}_i$ is the position operator associated with the 
	$i$-th particle space coordinates.
	As the operator in \eqref{eq:PartialF} commute with the relevant momentum 
	operator, the following commutation relations hold
	\begin{equation*}
		\big[\hat{l}^{\gamma}_i,\hat{\tau}^{\beta}_i\big]= 
		\mathrm{i}\hbar\epsilon_{{\gamma}{\beta}{\nu}}\hat{\tau}^{\nu}_i,		
	\end{equation*}
	where $\hat{\boldsymbol{l}}_i=\big(\hat{l}^{\nu}_i\big)_{\nu\in\mathbb{K}}$ is
	the corresponding orbital angular momentum operator.
	Therefore, the magnitude of the torque applied to the $i$-th particle, 
	which acquires an orbital angular momentum with quantum number $l_i$,
	reads $|\boldsymbol{\tau}_i|=\varepsilon\sqrt{\tau_i(\tau_i+1)}$, where 
	$\tau_i\equiv l_i$. 
	The equation giving the rate at which the corresponding orbital angular 
	momentum changes with time is analogous to the equation \eqref{eq:FpTime}. 
	It is written as
	\begin{equation*}
		\mathrm{d}_t\hat{\boldsymbol{\tau}}_i=
		\frac{\varepsilon}{\hbar}
		\mathrm{d}_t\hat{\boldsymbol{l}}_i
	\end{equation*}
	and shows us that any variation in the orbital angular momentum corresponds 
	to a change of the relevant torque by a constant of proportionality 
	$\hbar/\varepsilon$.
	
	In the relativistic case, the expression for the four-force operator follows
	from equation \eqref{eq:FourForce} such that one has
	\begin{equation*}
		\hat{\mathcal{F}}^\mu_i=\mathrm{i}\varepsilon\partial^\mu_i, 
	\end{equation*}
	where the derivative $\partial^\mu=(\partial_o,-\nabla)$.

\subsection{Uncertainty relations}\label{sec:un}
	
	With respect to the operator representations of the mechanical power and 
	impact force discussed in Sec. \ref{sec:xyz}, 
	the Heisenberg uncertainty principle applies.
	For all $i$ and $\nu\in\mathbb{K}$, one obtains the following inequalities
	\begin{subequations}\label{eq:Uncertain}
		\begin{equation}
			\triangle\nu_i\triangle F^{\nu}_i\geq\frac{\varepsilon}{2}
		\end{equation}
		and
		\begin{equation}
			\triangle t \triangle P_i\geq\frac{\varepsilon}{2}.
		\end{equation}
	\end{subequations}
	Therefore, gaining a knowledge for the position of a particle leads 
	to a lack of information for the applied force and hence the change 
	in its momentum.
	Furthermore, the greater the energy that a particle may transfer 
	the greater the uncertainty in power.

\subsection{Electromagnetic field example}\label{sec:elfield}
	
	Expressing the free-space electromagnetic field operators in terms of 
	\eqref{eq:PhaseFP}, 
	in addition to the Hamiltonian and total momentum operators of the 
	electromagnetic field, we obtain the power operator
	\begin{equation}\label{eq:ElP}
		\hat{P}=\sum_{\boldsymbol{k},\sigma}
		\varepsilon\omega_{\boldsymbol{k}}\hat{n}_{\boldsymbol{k},\sigma},
	\end{equation}
	where the number operator $\hat{n}_{\boldsymbol{k},\sigma}$ gives the number 
	of photons in a mode determined by the wave vector $\boldsymbol{k}$ and 
	the polarization index $\sigma$. The operator in \eqref{eq:ElP} 
	gives the maximum mechanical power generated when 
	creating the photons 
	$|(\boldsymbol{k},\sigma)^n,\ldots, (\boldsymbol{k}',\sigma')^{n'}\rangle$ 
	form the vacuum $|0\rangle$. 
	Respectively, the operator
	\begin{equation}\label{eq:ElF}
		\hat{F}=\sum_{\boldsymbol{k},\sigma}
		\varepsilon\boldsymbol{k}\hat{n}_{\boldsymbol{k},\sigma},
	\end{equation}
	describes the same processes in terms of applied force.
	In other words, it gives the applied force during the creation of the 
	aforementioned number of photons within the time-interval 
	$\hbar/\varepsilon$.


\section{Conclusion}\label{sec:conclusion}

	Studying the processes of energy and momentum transfer between interacting 
	particles in quantum mechanics, we obtain a lower limit for the duration 
	of these processes. Accordingly, the change in kinetic energy and momentum 
	of a particle is quantized. Introducing a quanta for the transfer of energy 
	and momentum leads to a unique representation of the 
	impulse-momentum theorem and hence relations for the respective 
	mechanical power, force and torque. 
	
	A key role in all derived relations plays a constant, denoted accordingly 
	by $\varepsilon$, that has a unit measure of energy and appears as 
	fundamental as the Planck's constant in the theory. 
	In particular, $\varepsilon$ is the proportionality constant between the 
	mechanical power of a quanta of interaction and the 
	frequency associated to that quanta, see equation 
	\eqref{eq:PErelation}.
	The presence of the named constant points out to the aforementioned lower
	limit for the rate at which the relevant energy and momentum transfer takes 
	place over time, see equations \eqref{eq:PF} and \eqref{eq:ImpulseQuantum}.
	
	In conclusion, studying the role of $\varepsilon$ in physics more closer,
	we definitely have to test its uniqueness and make efforts in determining 
	its value.

	
	\begin{acknowledgments}
		The author is indebted to H. Chamati, J. Boradjiev and S. Varbev for the very useful 
		discussions and to 
		S. Kazakova, D. Dakova and E. Pisanova 
		for their guidance in writing down the pilot idea shortly revisited 
		in the present paper.
		This work was supported by the Bulgarian Ministry of Education and Science, 
		National program ``Yong science and postdoctoral 
		researchers'' approved by DCM 577, 17.08.2018.
	\end{acknowledgments}




%

\end{document}